\documentstyle[12pt,aasms4]{article}
\begin{document}

\title{FUSE Spectra of the Black Hole Binary LMC X-3\footnote{Based on
observations made with the NASA-CNES-CSA Far Ultraviolet Spectroscopic
Explorer. FUSE is operated for NASA by Johns Hopkins University under 
NASA contract NAS 5-3298}}

\author{J.B. Hutchings, K. Winter}
\affil{Herzberg Institute of Astrophysics, National Research Council of
Canada, \\ 5071 W.\ Saanich Rd., Victoria, B.C. V9E 2E7, Canada \\ email
john.hutchings@nrc-cnrc.gc.ca} 

\author{A.P. Cowley, P.C. Schmidtke}  
\affil{Department of Physics \& Astronomy, Arizona State University,
Tempe, AZ, 85287-1504 }

\and

\author{D. Crampton}
\affil{Herzberg Institute of Astrophysics, National Research Council of 
Canada, \\ 5071 W.\ Saanich Rd., Victoria, B.C. V9E 2E7, Canada}

\begin{abstract}

Far-ultraviolet spectra of LMC X-3 were taken covering photometric phases
0.47 to 0.74 in the 1.7-day orbital period of the black-hole binary (phase
zero being superior conjunction of the X-ray source).  The continuum is
faint and flat, but appears to vary significantly during the observations.
Concurrent $RXTE$/ASM observations show the system was in its most
luminous X-ray state during the FUSE observations.  The FUV spectrum
contains strong terrestrial airglow emission lines, while the only stellar
lines clearly present are emissions from the O~VI resonance doublet. 
Their flux does not change significantly during the FUSE observations. 
These lines are modelled as two asymmetrical profiles, including the local
ISM absorptions due to C~II and possibly O~VI.  Velocity variations of
O~VI emission are consistent with the orbital velocity of the black hole
and provide a new constraint on its mass. 

\end{abstract}


\section{Introduction}

LMC X-3 is one of the brightest X-ray sources in the Large Magellanic
Cloud, with L$_X>10^{38}$ erg s$^{-1}$.  It shows both high/soft and
low/hard X-ray states.  Spectroscopic observations of the B3~V star
revealed an orbital period of $\sim1.7$ days and an unseen massive
companion which was interpreted as a $\sim10$M$_{\odot}$ black hole
(Cowley et al. 1983).  Subsequent optical photometry (van der Klis et al.
1985; Kuiper et al. 1988) showed ellipsoidal variations
($\Delta$m$\sim0.2$ mag) of the B star, which improved the ephemeris and
also indicated a large mass for the secondary star.  Ultraviolet spectra
taken with HST showed a moderate strength N~V, 1240\AA, emission line and
weak emissions of C~IV, 1550\AA, and He~II, 1640\AA\ (Cowley et al. 1994).
These lines appeared to be formed in the accretion disk surrounding the
black hole, but only two spectra were obtained and it was not possible to
determine their velocity amplitude.  Given the strength of N~V, we
expected the resonance lines of O~VI to be detectable with FUSE spectra. 
The properties of these lines might provide further information about the
accretion disk and motion of the black hole. 

\section{FUSE Observations}

LMC X-3 was observed through 8 consecutive FUSE orbits, yielding spectra
centered at binary phases 0.47 through 0.74, thus covering about a quarter
of the orbit beginning slightly before the inferior conjunction of the
X-ray source.  The observations were taken through the large science
aperture and are summarized in Table 1.  Spectra were extracted using the
standard FUSE pipeline procedure. 

The continuum is weak across the FUSE bandpass, with flux roughly flat at
$8\times10^{-15}$ erg s$^{-1}$ cm$^{-2}$ \AA$^{-1}$.  One FUSE channel
(SiC A) shows no continuum, which probably means the alignment was not
good for that telescope.  All channels show a rich emission-line spectrum
of H~I and O~I, which arises from terrestrial airglow.  The zero radial
velocity of these emission lines confirms this origin.  The strength of
these lines changes from day to night (O~I drops more than H~I at night),
so in Table 1 we have indicated what percentage of each observation was
taken during nighttime.  The observations were taken with a single
pointing with re-acquisitions after earth-occultations, so the different
FUSE channels should have had stable alignment throughout the
observations. 

The presence of the airglow lines makes it difficult to detect many of the
principal high ionization lines that may be present in the LMC X-3
spectrum, such as C~III, 977\AA, N~III, 992\AA, and He~II, 1085\AA.
However, if present, they are very weak.  There is a possible weak broad
feature at C~III, 1175\AA, and a possible feature at $\sim$1000\AA\ which
is clear of airglow but has no obvious identification.  However, the
principal feature of the spectrum is the O~VI doublet, which is stronger
than any of the other lines and mostly not overlaid by airglow emission. 
The far ultraviolet spectrum of LMC X-3 is shown in Figure 1.

\section{Concurrent X-ray Data}

We have also extracted the $RXTE$ All Sky Monitor (ASM) data from the
publicly available web site.  These data are shown in Figure 2, with the
time of the FUSE observations indicated.  LMC X-3 is known to vary
considerably in its X-ray flux, and the figure shows that the FUSE data
were obtained when the source was in its brightest X-ray phase.  We note
that early observations suggested the occurence of the high/low states
might be periodic (Cowley et al. 1991), but this has been demonstrated not
to be the case (e.g. Wilms et al. 2001; Brocksopp, Groot, \& Wilms 2001). 
These authors find that the optical and X-ray light curves are correlated,
with the X-rays slightly lagging in time, implying a variable
mass-accretion rate model rather than disk precession. 

\section{FUSE Measurements}

Table 1 gives the measures of the continuum and O~VI line flux from the
FUSE spactra.  The continuum was estimated from the LiF channels 1a, 2a,
and 2b (1b has the `worm' grid wire shadow that masks some flux), after
removing the airglow emission lines.  The variations are seen in all three
channels and so are considered real.  The absolute level of the continuum
is less certain because of the difficulty in extracting background and
detector systematics at this low flux level.  The general nature of the
changes are a $\sim$20\% rise in spectrum through phase 0.64, followed by
a steady drop by a factor $\sim2$ over the last three spectra.  These
relative changes in the continuum have scatter of less than 20\% after the
mean levels have been normalized. 

The O~VI line flux was measured from all four FUSE channels that cover
this wavelength range, with greater weight on the LiF channels.  There is
considerable noise and strong C~II absorption within the profile, so the
numbers given have standard deviations of about 15\%.  Thus, there is no
significant variation in the O~VI flux, although there is a suggestion
that the line strength rises and then falls in phase with the continuum. 

\section{Discussion}

The continuum changes occur over binary phases where the accreting black
hole moves from inferior conjunction to quadrature.  At these phases, a
disk thickening at the impact point of the mass-transfer stream would move
from illuminated to partially covering the disk, which could account for
the observed drop in far-UV flux.  However, there may also be non-phased
changes on this timescale.  The only way to clarify the matter would be to
obtain observations that cover additional binary phases, as well as
several orbital cycles.  The small O~VI flux changes may occur in the
unocculted central parts of the accretion disk. 

The optical ellipsoidal light curve has an amplitude of $\sim$0.2 mag (van
der Klis et al. 1985), while a small amount of data in the UV from HST
shows a similar range.  The orbital X-ray flux changes are smaller (Boyd,
Smale, \& Dolan 2001).  Thus, the indicated large flux range reported here
in the FUV has not been seen at other wavelengths. 

If O~VI emission arises in the inner disk, it should show the orbital
motion of the black hole.  Before looking for line shifts, we need to
understand the observations in more detail.  Figures 3 and 4 display the 
O~VI region in different ways and show that there are strong absorptions at
1036-7\AA\ from C~II.  These are close to their rest wavelengths, so they
are identified as Galactic ISM features.  Figure 4 also shows O~VI
absorptions at their rest wavelengths (one is blended with C~II, but the
relative line strengths are consistent with that).  Thus, in looking for
line shifts from LMC X-3, we need to ignore these local absorption
components.  (It is of interest to note the apparent narrow O~VI absorber
at rest velocity.  It is unlikely to arise near LMC X-3 in an outflow at
exactly the velocity of LMC X-3 with respect to us, so it must arise in
the hot ISM of the Galaxy.) 

In Figure 4, we show a possible way to model the O~VI line emission.  We
have used just the LiF1a channel data for this, as it has the highest S/N
and has the most reliable wavelength scale, being the guide channel.  The
emission is assumed to have an aymmetrical profile defined by the long
wavelength tail and the shortward cutoff of the whole feature.  We further
assume the two O~VI lines have identical profiles, that their strengths
have roughly the ratio of 2 expected for optically thin emission, and that
the peaks are centered at the systemic velocity of LMC X-3 (+310 km
s$^{-1}$).  The result, shown in Figure 4, fits the overall profile well.
It is possible to fit the data with symmetrical profiles too, but this
would require a different identification for the emission longward of
1042\AA, and none is obvious. 

To look for velocity changes, we summed spectra into two phase bins, as
illustrated in Figure 3, since the spectra are too noisy to measure
individually.  The bins are centered at phases 0.53 and 0.70.  Velocity
shifts between the two phase bins were measured by: a) fitting of profiles
as shown in Figure 4; b) cross-correlation between them and the overall
mean, both with the C~II absorptions in and edited out; and c) hand
overlay of the plotted profiles, filtering the absorbers out visually.  We
also worked with both the LiF1a alone, and sum of all channel spectra.  In
all measurements, we found the 0.53-phase bin to have a more negative
velocity than the 0.70-phase bin, with the velocity difference being in
the range 100-150 km s$^{-1}$.  When we fit a sine curve to this
difference, imposing the phasing for expected orbital motion of the black
hole, we find an orbital semiamplitude, K$_{bh}$, in the range of 130-200
km s$^{-1}$.  Cowley et al.\ (1983) found K$_B$=235 km s$^{-1}$ from
measuring the B star's velocities.  Hence, the FUSE results suggest the
black hole may be more massive than the optically visible star.  The
implied mass ratio lies in the range 0.55$-$0.83.  Adopting the highest of
these values, we derive minimum masses of 15 and 13 M$_{\odot}$ for the
black hole and B star, respectively. 

We note that Soria et al.\ (2001) suggest that the velocity amplitude
derived by Cowley et al.\ may be overestimated by $\sim15$\% due to
possible line weakening on the X-ray heated face of the B star.  However,
this scenario would require a significant modulation of the spectrum with
phase, which is not observed.  A plot of the equivalent widths of both the
hydrogen and He~I lines versus phase (using those tabulated by Cowley et al.,
and orbital phases computed using the ephemeris of van der Klis et al.) 
shows that there is no orbital trend in the strength of these lines.  As 
noted by Cowley et al., all lines are weakened compared to a normal B star,
presumably because of the additional continuum from the accretion disk.
However, the ratio of H to He~I line strengths is similar to that in
normal single B stars.  On the other hand, it is possible that the B-star
velocity semiamplitude might instead be \underbar{underestimated} if the
heating effect causes one to observe more spectral light from the B star's
inner hemisphere. 

Our FUSE measurement from averaged spectra of only two emission velocites,
which are thought to arise in the black hole's accretion disk, is far from
a conclusive determination of the mass ratio in LMC X-3.  However, it does
suggest that full orbital observations of UV emission lines could help to
define the range of possible masses in this system.  Thus, while the FUSE
data are quite limited, they are consistent with the O~VI emission
originating in the inner parts of the accretion disk and provide a direct
estimate of the mass ratio for the first time.  Further orbital coverage
by FUSE would provide much more definitive mass estimates. 

\acknowledgments The authors acknowledge the use of the $RXTE$-ASM data
from the web site.  In addition, APC thanks NASA for partial support of
this research. 

\clearpage

\begin{deluxetable}{cccccc}
\tablenum{1}
%\tablecolumns{6}
%\footnotesize
\tablecaption{FUSE Observations of LMC X-3}
\tablehead{
\colhead{MJD} &
\colhead{Exp time} &
\colhead{\% Obs} &
\colhead{$\Phi_{phot}$\tablenotemark{a}} &
\colhead{Continuum} &
\colhead{O~VI emission}\\ 
\colhead{(mid-exp)} &
\colhead{(sec)} &
\colhead{at night} & &
\colhead{(erg s$^{-1}$ cm$^{-2}$ \AA$^{-1}$) } &
\colhead{ (erg s$^{-1}$ cm$^{-2}$) } 
}

\startdata
52228.742 &2929 &69 &0.47 &6.7$\times10^{-15}$ &8.1$\times10^{-14}$ \nl
52228.813 &2673 &64 &0.52 &7.1$\times10^{-15}$ &7.1$\times10^{-14}$ \nl
52228.884 &2467 &60 &0.56 &7.6$\times10^{-15}$ &9.5$\times10^{-14}$ \nl
52228.956 &2262 &53 &0.60 &7.2$\times10^{-15}$ &8.8$\times10^{-14}$ \nl
52229.027 &2153 &46 &0.64 &9.1$\times10^{-15}$ &8.8$\times10^{-14}$ \nl
52229.108 &4059 &24 &0.69 &7.0$\times10^{-15}$ &9.1$\times10^{-14}$ \nl
52229.154 &3942 &54 &0.72 &5.5$\times10^{-15}$ &8.1$\times10^{-14}$ \nl
52229.200 &3942 &28 &0.74 &3.9$\times10^{-15}$ &7.5$\times10^{-14}$ \nl
\enddata

\tablenotetext{a}{ Photometric ephemeris: HJD 2445278.005 + 1.70479E (MJD 
45277.505), where $\Phi$=0 is superior conjunction of black hole (from
van der Klis et al.\ 1985) }

\end{deluxetable}

\clearpage

\begin{figure}
%fig 1
\caption{Overall spectrum of LMC X-3, from the best FUSE channels.  The
shorter wavelengths (top panel) have more smoothing for easier visibility. 
The sharp emission lines are all airglow, and they are broadened by the
smoothing.  The positions of the high ionization emission lines seen in
accretion disk systems are marked (at LMC shifted wavelengths).  Only O~VI
is definitely present.  Both LiF channels are plotted in this region to
show the detailed agreement.} 
\end{figure}

\begin{figure}
%fig 2
\caption{$RXTE$ All Sky Monitor X-ray data for LMC X-3 covering $\sim$8 
years of observations.  The date of the FUSE observations presented in 
this paper is marked.}
\end{figure}

\begin{figure}
%fig 3
\caption{ O~VI emission region, overall and in two binary phase bins.  The
lower panel shows the difference between the two phase-binned spectra.  The
dotted lines indicate the regions where airglow emission has been edited
out.  These spectra are derived from all FUSE channels.} 
\end{figure}

\begin{figure}
%fig 4
\caption{Mean O~VI LiF1a profile with fitted profiles of the doublet (shown by
the lighter weight lines - see text).  Dashed line shows the residual spectrum
after subtraction of model profiles.  The positions of local absorbers of
C~II and O~VI are marked.} 
\end{figure}


\begin{references}

\reference{} Boyd, P.T., Smale, A.P., \& Dolan, J.F. 2001, ApJ, 555, 822

\reference{} Brocksopp, C., Groot, P.J., \& Wilms, J. 2001, MNRAS, 328, 
139

\reference{} Cowley, A.P., Crampton, D., Hutchings, J.B., Remillard, R.,
\& Penfold, J.E. 1983, ApJ, 272, 118 

\reference{} Cowley, A.P., Schmidtke, P.C., Ebisawa, K., Makino, F., 
Remillard, R.A., Crampton, D., Hutchings, J.B., Kitamoto, S., \& Treves, 
A. 1991, ApJ, 381, 526

\reference{} Cowley, A.P., Schmidtke, P.C., Hutchings, J.B., \&
Crampton, D. 1994, \apj, 429, 826

\reference{} Kuiper, L., van Paradijs, J., \& van der Klis, M. 1988, \aap, 
203, 79

\reference{} Soria, R., Wu, K., Page, M.J., \& Sakelliou, I. 2001, A\&A,
365, L273 

\reference{} van der Klis, M., Clausen, J.V., Jensen, K., Tjemkes, S., \&
van Paradijs, J. 1985, A\&A, 151, 322 

\reference{} Wilms, J., Nowak, M.A., Pottschmidt, K., Heindl, W.A., Dove, 
J.B., \& Begelman, M.C. 2001, MNRAS, 320, 327

\end{references}
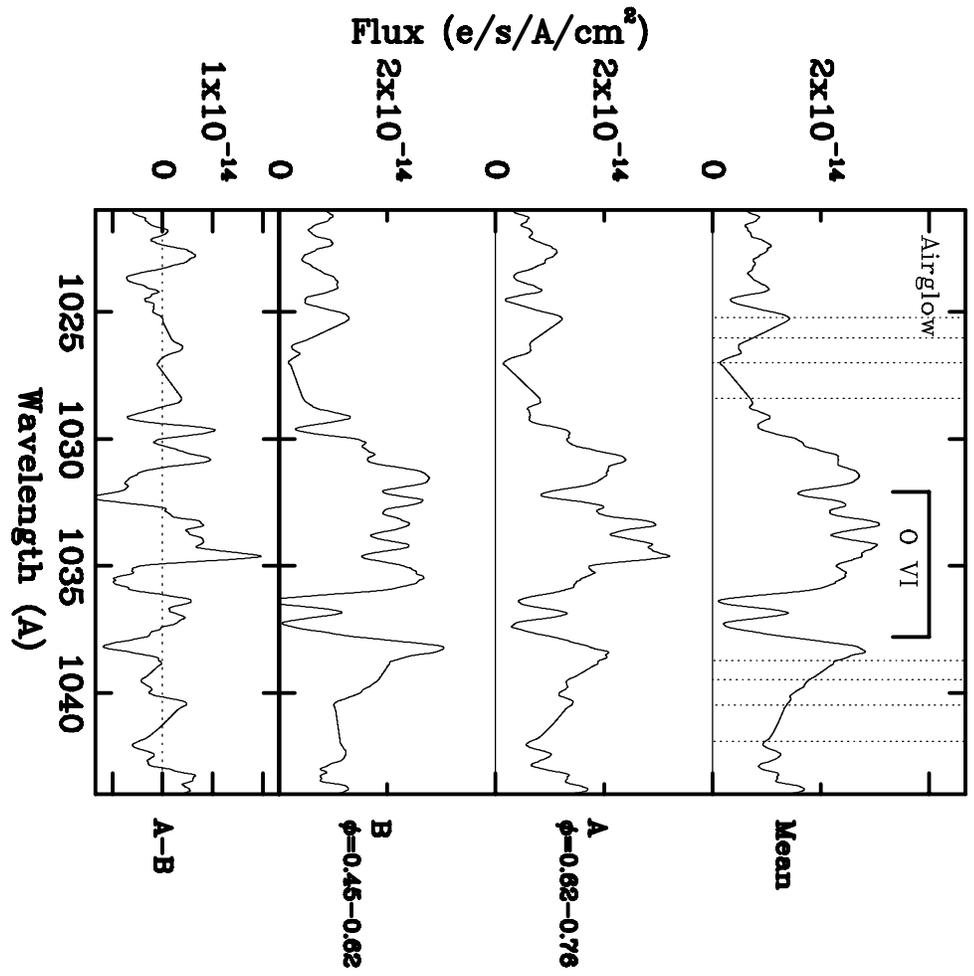
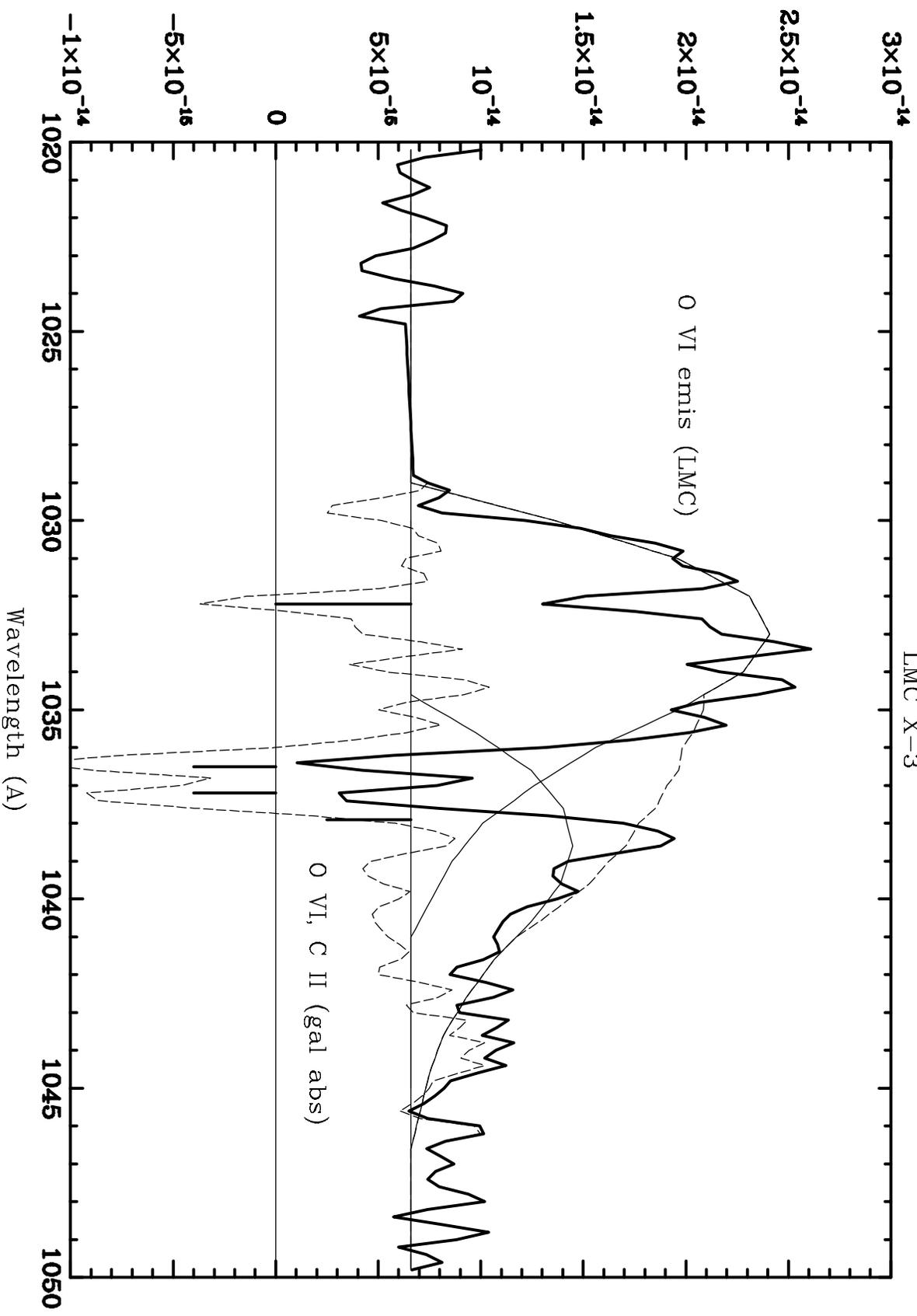
\end{document}